\documentclass[onecolumn,amsmath,amssymb,pre]{revtex4-2}

\usepackage{graphicx}
\usepackage[dvipsnames]{xcolor}
\usepackage{dcolumn}
\usepackage{amsmath}
\usepackage{subfigure}

\usepackage{sidecap}
\usepackage{dsfont}

\usepackage{mathrsfs}
\usepackage{amsfonts}
\usepackage{indentfirst}
\usepackage{bm}
\usepackage{tikz}

\usepackage[colorlinks,urlcolor=NavyBlue,citecolor=TealBlue]{hyperref}

\newcommand{\be}{\begin{equation}}
\newcommand{\ee}{\end{equation}}
\newcommand{\bey}{\begin{eqnarray}}
\newcommand{\eey}{\end{eqnarray}}
\newcommand{\bw}{\begin{widetext}}
\newcommand{\ew}{\end{widetext}}

\newcommand{\ra}{\rangle}
\newcommand{\la}{\langle}

\newcommand{\ba}{\begin{array}}
\newcommand{\ea}{\end{array}}
\newcommand{\bi}{\begin{itemize}}
\newcommand{\ei}{\end{itemize}}
\newcommand{\bem}{\begin{enumerate}}
\newcommand{\eem}{\end{enumerate}}

\begin{document}

\title{Finite time quantum-classical correspondence 
in quantum chaotic systems}

\author{Qian Wang} 
\affiliation{CAMTP-Center for Applied Mathematics and Theoretical Physics, University of Maribor, 
Mladinska 3, SI-2000 Maribor, Slovenia, and \\
Department of Physics, Zhejiang Normal University, Jinhua 321004, China}
\author{Marko Robnik} 
\affiliation{CAMTP-Center for Applied Mathematics and Theoretical Physics, University of Maribor, 
Mladinska 3, SI-2000 Maribor, Slovenia}

\begin{abstract}

Although the importance of the quantum-classical correspondence has been
recognized in numerous studies of quantum chaos, whether it still holds for finite time 
dynamics remains less known.  
We address this question in this work by performing a detailed analysis of how 
the quantum chaotic measure relates to the chaoticity of the finite time classical trajectories.
A good correspondence between them has been revealed 
in both time dependent and many-body systems. 
In particular, we show that the dependence of the quantum 
chaotic measure on the chaoticity of finite time trajectories 
can be well captured by a function that is independent of the system.
This strongly implies the universal validity of the 
finite time quantum-classical correspondence.
Our findings provide a deeper understanding of the 
quantum-classical correspondence and highlight
the role of time for studying quantum ergodicity.

\end{abstract}

\date{\today}

\maketitle

\section{Introduction}

Since the beginning of the quantum theory, the quantum-classical correspondence 
principle has attracted countless investigations and remains an interesting
topic in current researches.
Roughly speaking, it states that there should be an agreement 
between the quantum and classical predictions in certain classical limit.
There are several formulations of the quantum-classical correspondence principle
\cite{Bohr1920,Nielsen2013,Wilkie1997a,Wilkie1997b}.
Among them, the most famous one is the Bohr's corresponding principle 
\cite{Bohr1920,Nielsen2013}, which plays a crucial 
role for interpretation of quantum mechanics.
Although the quantum-classical correspondence principle 
has been verified in abundant literature, see e.~g.~Refs. 
\cite{Keller1958,Kryvohuz2005,Gutzwiller2013,Jarzynski2015,
Kumari2018,WangJ2020,Arranz2021,Vijaywargia2024},
it meets several difficulties in the studies of quantum chaos. 

On the one hand, it is known that the energy spectrum 
of any bounded quantum system is discrete.
As a result, the quantum motion should always be governed by the regular dynamics, 
rather than the chaotic one, which appears for the continuous spectrum.
However, this is in conflict with the observed classical dynamical chaos \cite{Gaspard2001}
and indicates the failure of the traditional quantum-classical correspondence. 
In fact, the chaotic behavior exhibited in the system with discrete 
spectrum requires us to modify the quantum-classical correspondence 
principle by considering the finite time behavior. 
On the other hand, different from the classical systems, the Heisenberg
uncertainty relation leads to a finite resolvable limit 
in the phase space of quantum systems.  
This further suppresses the quantum chaotic motion 
to agree with its classical counterpart \cite{Toda1989}.    
Moreover, the quantum interference effect among 
the phase space structures of quantum systems 
gives rise to localization phenomena 
even in strongly chaotic systems. 
Consequently, the quantum chaotic systems with well-defined classical limit 
could allow their spectral statistics to exhibit significant deviations  
from the expected Wigner-Dyson statistics 
\cite{Borgonovi1996,Neilson1998,Batisic2013,WangG2022,Lozej2022}.
Additionally, the extremely slow diffusion dynamics in classical systems
\cite{Altmann2005,WangJ2014,HuangL2017,Lozej2021,Zahradova2022}
provides an alternative mechanism that leads to the breakdown 
of the quantum-classical correspondence in the finite time.  
                
All these facts together raise an intriguing question of whether 
the quantum-classical correspondence still holds for the finite time motion.  
A very recent work performed by Casati and coworkers has explored 
this question in several billiards \cite{HCasati2024}. 
By associating the finite time classical trajectories with a certain energy shell of 
the quantum system, a well-defined correspondence between
the quantum and classical chaotic measures has been established in their studies.
This provides an evidence of the validity of the quantum-classical correspondence 
in the finite time dynamics.
However, a general conclusion remains elusive, 
since the billiards are the single body systems.
Hence, it is necessary to investigate the finite time quantum-classical correspondence
in different quantum systems, particularly the many-body systems, 
to enhance the conclusion in above mentioned study.

In this work, utilizing the concepts proposed in Ref.~\cite{HCasati2024}, 
we carry out a detailed investigation of the finite time 
quantum-classical correspondence in two 
quantum systems that are distinct from the billiards.
The first one is the kicked top model, 
a prototypical model for studying of quantum chaos \cite{Haake2019}.  
Although the kicked top model is also a single particle system, it has
time-dependent Hamiltonian that distinguishes it from the billiards. 
The second system is a many-body system, 
namely the well-known Dicke model \cite{Dicke1954},
which has a smooth classical Hamiltonian limit,
in general of the mixed type classical dynamics. 
Different from both billiards and kicked top model,
the classical dynamics of the Dicke model is governed 
by differential equations, rather than the mapping.

By employing the Poincar\'e section, 
we show how to define the classical chaotic measure for 
a finite time classical trajectory.
We demonstrate that the classical chaoticity quantified by finite time trajectories 
exhibits a good agreement with the quantum chaoticity.  
In particular, we find that the dependence between the 
quantum and classical chaotic measures of totally different systems 
can be captured by the same function.
This remarkable finding not only extends the 
previous results in Ref.~\cite{HCasati2024} to more general systems, 
but also promotes our understanding of 
the quantum-classical correspondence principle. 
In addition, it further implies the existence of a universal relationship
between the quantum and finite time classical chaotic measures.

The rest of this article is organized as follows.
In Sec.~\ref{Secd}, we first introduce the kicked 
top model and then report on our numerical results 
of the finite time quantum-classical correspondence.
The Sec.~\ref{third} is devoted to the study of a
different model, the Dicke model. 
Here, we show that the numerical results obtained from 
the Dicke model are similar to the kicked top model, 
suggesting the universality of the finite time 
quantum-classical correspondence. 
We finally summarize our findings in Sec.~\ref{fourth} 
with several remarks.

 \begin{figure}
  \includegraphics[width=\textwidth]{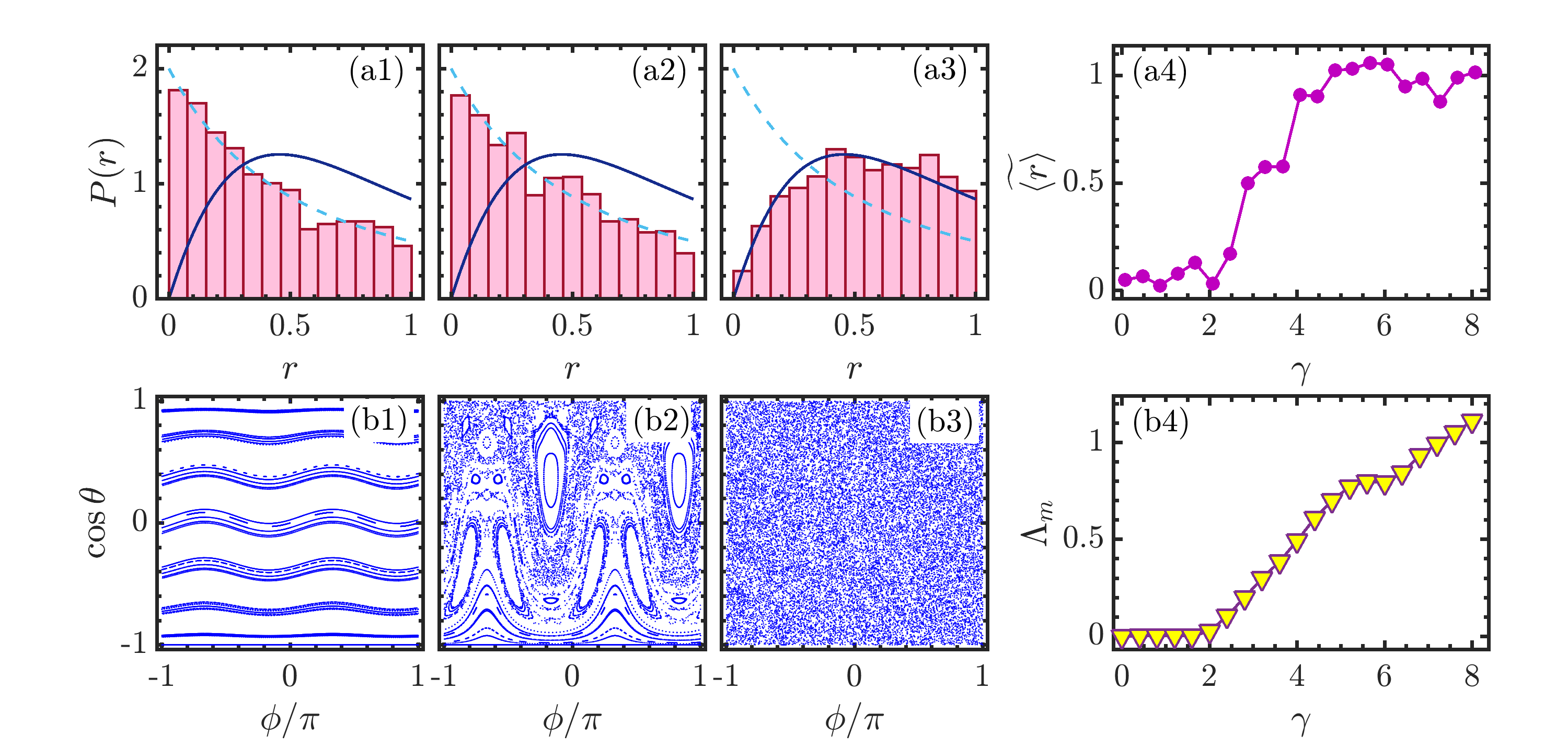}
  \caption{(a1)-(a3): Level spacing ratio probability distribution in Eq.~(\ref{GapR}) 
  of the KT model for $\gamma=0.2$ (a1),
  $\gamma=2.3$ (a2), and $\gamma=7$ (a3) with $j=2000$.
  The dashed and solid curves in each panel denote $P_P(r)$ 
  and $P_{WD}$ in (\ref{PWD}), respectively.
  (a4) Averaged ratio $\widetilde{\la r\ra}$ [cf.~Eq.~(\ref{Ravr})] of the KT model 
  as a function of $\gamma$ for $j=2000$. 
  (b1)-(b3): Classical phase space portraits of the KT model
  for $90$ random initial conditions with 
  $\gamma=0.2, 2.3$, and $\kappa=7$ [from (b1) to (b3)].
  Each initial condition has been evolved for $300$ kicks. 
  (b4) Phase space averaged Lyapunov exponent $\Lambda_m$ (\ref{AvgLp})
  of the KT model as a function of $\gamma$.
  Here, $\Lambda_m$ is numerically obtained by averaging over $2\times10^4$ initial conditions, 
  each evolved for $1\times10^5$ kicks. 
  Other parameter: $\beta=\pi/3$.}
  \label{QCchaosKT}
 \end{figure}

\section{Kicked top model} \label{Secd}

The kicked top (KT) model \cite{Haake1987,Haake2019}, 
firstly introduced in the context of quantum chaos 
\cite{Ariano1992,Fox1994,Bandy2004,Ghose2008,Lombardi2011,Dogra2019,
Herrmann2020,Olsacher2022,QianWR2023a,AnandG2024}, 
describes the periodic kicking on a precessing top with 
angular momentum $\mathbf{J}=(J_x,J_y,J_z)$.
This model has been extensively studied in many areas of physics
\cite{Strzys2008,Bastidas2014,Bhosale2018,Sieberer2019,Mondal2021,
Munoz2021,YChao2021,Alonso2022,QianWR2023b,Anand2024,Passarelli2024} 
and can be realized in different experimental platforms 
\cite{Chaudhury2009,Krithika2019,Meier2019}.
The Hamiltonian of the KT model can be written as
\be \label{KTH}
   H_{KT}=\beta J_z+\frac{\gamma}{2j}J_x^2\sum_{n=-\infty}^{+\infty}\delta(t-n),
\ee
where $\beta$ is the angular frequency of the precession around $z$ axis,
$\gamma$ denotes the strength of the $\delta$ kicks with unit period, 
and $j$ quantifies the total magnitude of $\mathbf{J}$, so that $\mathbf{J}^2=j(j+1)$. 
Here, we set $\hbar=1$ throughout this work.

The conservation of $\mathbf{J}^2$ in the KT model  allows us to work 
in the Dicke basis $|j,m\ra$, defined by $j_z|j,m\ra=m|j,m\ra$ and 
$\mathbf{J}^2|j,m\ra=j(j+1)|j,m\ra$ with $-j\leq m\leq j$,
and the dimension of the Hilbert space is $\mathcal{D}=2j+1$.
Moreover, the Hamiltonian (\ref{KTH}) also conserves the parity operator $\Pi=(-1)^{(j+J_z)}$.
Consequently, the Hilbert space can be further split into even- and odd-parity subspaces 
with dimensions $\mathcal{D}_e=j+1$ and $\mathcal{D}_o=j$ for even $j$.
We focus on the even-parity subspace in this work and consider $j$ even.

It is known that the KT model exhibits a transition to chaos 
with increasing the kick strength \cite{Haake2019,QianWR2023a,WangX2004,Piga2019,Lerose2020}.
The onset of chaos in the KT model can be analyzed through the spectral properties of
the Floquet operator, which can be written as
\be \label{Floquet}
   \mathcal{F}=\exp\left(-i\frac{\gamma}{2j}J_x^2\right)\exp(-i\beta J_z).
\ee 
By numerically diagonalizing $\mathcal{F}$ in the Dicke basis, 
we obtain its eigenvalues $\{\alpha_k\}\in[-\pi,\pi)$, also known 
as the quasienergies of the KT model.
Then, we consider the level spacing ratios \cite{Oganesyan2007}, defined as
\be \label{GapR}
   r_k=\mathrm{min}\left(\delta_k,\frac{1}{\delta_k}\right),
\ee
where $\delta_k=d_{k+1}/d_k$ with $d_{k+1}=\alpha_{k+1}-\alpha_k$ being the spacing 
between two nearest quasienergies. 
The presence of chaos can be revealed by the 
probability distribution of $r_k$, denoted by $P(r)$. 
It has been verified that for the regular and 
fully chaotic systems $P(r)$ are, 
respectively, captured by the Poisson and 
Wigner-Dyson distributions \cite{Atas2013a,Atas2013b,Giraud2022},
\be \label{PWD}
   P_{P}=\frac{2}{(1+r)^2},\quad
   P_{WD}=\frac{27}{4}\frac{r+r^2}{(1+r+r^2)^{5/2}}.
\ee 

In Figs.~\ref{QCchaosKT}(a1)-\ref{QCchaosKT}(a3), 
we plot how $P(r)$ varies as $\gamma$ is increased.
The crossover from $P_P(r)$ to $P_{WD}(r)$ with 
increasing $\gamma$ is clearly visible, 
indicating the occurrence of chaos at larger values of $\gamma$. 
Further confirmation of the transition to chaos is provided 
by the rescaled average ratio, defined by \cite{QianWR2023a}
\be\label{Ravr}
  \widetilde{\la r\ra}=\frac{|\la r\ra-\la r\ra_P|}{\la r\ra_{WD}-\la r\ra_P}.
\ee
Here, $\la r\ra$ is the average of $r_k$, $\la r\ra_P=\int rP_P(r)dr\approx0.39$, and 
$\la r\ra_{WD}=\int rP_{WD}(r)dr\approx0.53$ \cite{Atas2013a}. 
The results in Figs.~\ref{QCchaosKT}(a1)-\ref{QCchaosKT}(a3) suggest that  
$\widetilde{\la r\ra}$ keeps its value around the
zero for smaller values of $\gamma$, while it 
saturates near $\widetilde{\la r\ra}\approx1$ when $\gamma$ is sufficient large.  
This is displayed in Fig.~\ref{QCchaosKT}(a4), where the dependence 
of $\widetilde{\la r\ra}$ on $\gamma$ is shown.

 \begin{figure}
  \includegraphics[width=\textwidth]{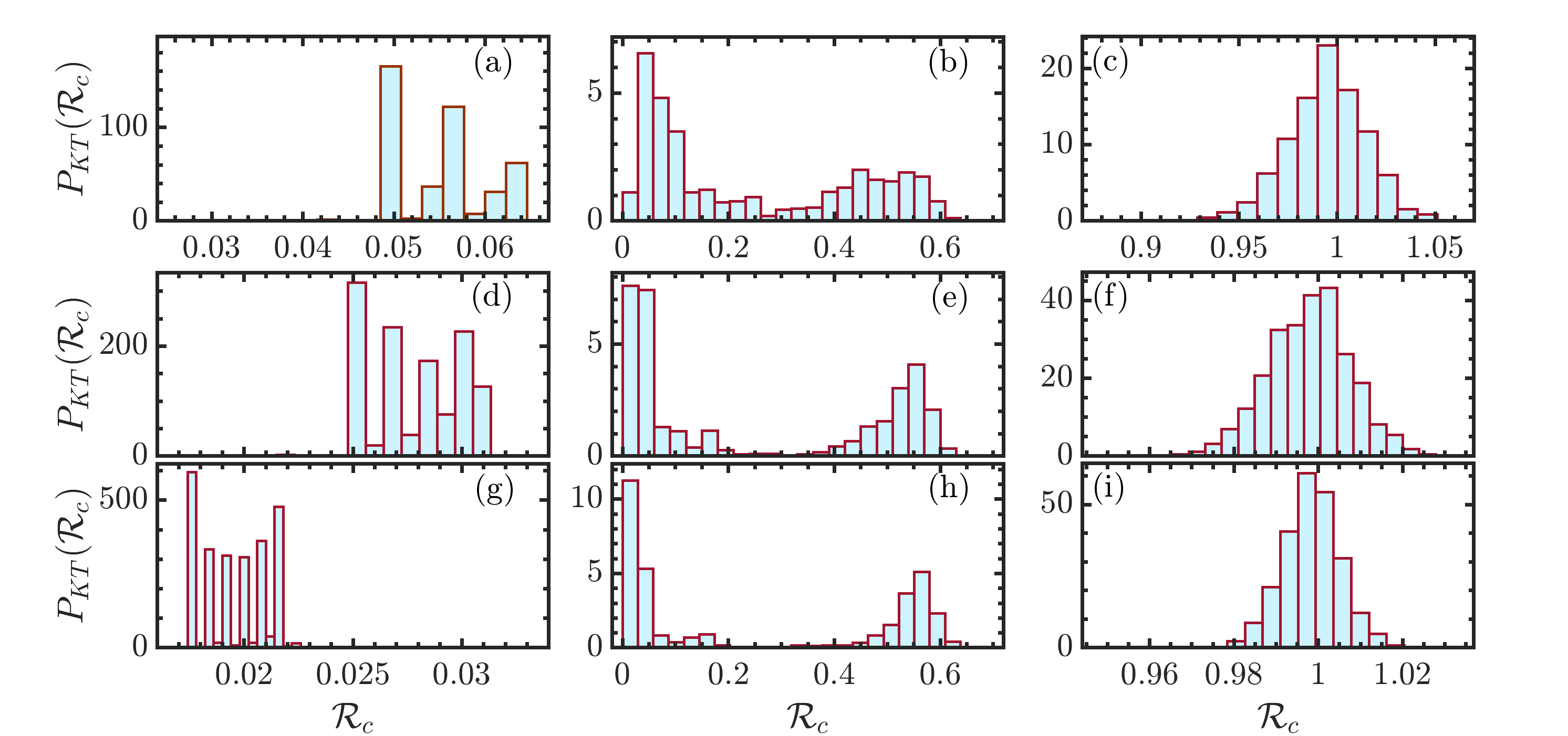}
  \caption{(a)-(c): Probability distribution, $P_{KT}(\mathcal{R}_c)$, of $\mathcal{R}_c$ in Eq.~(\ref{KTRc}) 
  for $\gamma=0.2$ (a), $\gamma=2.3$ (b) and $\gamma=7$ (c) with $N_k=1000$.
  (d)-(f): $P_{KT}(\mathcal{R}_c)$ for the same values of $\gamma$ as in panels (a)-(c) with $N_k=4000$. 
  (g)-(i): $P_{KT}(\mathcal{R}_c)$ for the same $\gamma$ values as in panels (a)-(c) with $N_k=8000$. 
  Here, an ensemble of $1600$ initial conditions has been employed to calculate 
  the probability distribution $P_{KT}(\mathcal{R}_c)$.
  Other parameter: $\beta=\pi/3$.}
  \label{PdfKT}
 \end{figure}

The transition to chaos in (\ref{KTH}) can be considered as a quantum manifestation of
the chaotic motion in its classical counterpart. 
The classical KT model is described by the following classical map \cite{QianWR2023a}
\begin{align} \label{CKT}
\begin{bmatrix}
 X_{n+1} \\
 Y_{n+1} \\
 Z_{n+1}
 \end{bmatrix}=
 \begin{pmatrix}
  \cos\beta &  -\sin\beta  &  0 \\
  \sin\beta\cos\Theta_n  &  \cos\beta\cos\Theta_n  &  -\sin\Theta_n  \\
  \sin\beta\sin\Theta_n   &  \sin\beta\sin\Theta_n    &  \cos\Theta_n  
 \end{pmatrix}
 \begin{bmatrix}
  X_n \\
  Y_n \\
  Z_n
 \end{bmatrix},
 \end{align} 
where $\Theta_n=\gamma(X_n\cos\beta-Y_n\sin\beta)$. 
The classical vector $\mathbf{X}=(X,Y,Z)$ satisfies $|\mathbf{X}|^2=X^2+Y^2+Z^2=1$.
We thus parametrize $(X,Y,Z)$ as $X=\sin\theta\cos\phi, Y=\sin\theta\sin\phi$ 
and $Z=\cos\theta$, with $\theta,\phi$ being the azimuthal and polar angles, respectively. 
The classical phase space is characterized by canonical 
variables $\phi=\arctan(Y/X)$ and $\cos\theta$.  

The classical phase portraits for several kick strengths 
are plotted in Figs.~\ref{QCchaosKT}(b1)-\ref{QCchaosKT}(b3).
As $\gamma$ is increased, the phase space evolves from occupied by 
regular orbits ($\gamma=0.2$) to the celebrated 
Kolmogorov-Arnold-Moser scenario ($\gamma=2.3$), and to the almost entirely
covered by chaotic trajectories, with invisible regular islands ($\gamma=7$). 
These observed features can be quantitatively captured by 
the phase space averaged Lyapunov exponent, defined by
\be \label{AvgLp}
   \Lambda_m=\frac{1}{4\pi}\int \lambda_m \sin\theta d\theta d\phi.
\ee
Here, $\lambda_m$ is the largest Lyapunov exponent 
of the classical map (\ref{CKT}) and it can be calculated as \cite{Ariano1992,Parker2012}
$\lambda_m=\ln[\lim_{n\to\infty}(\tau_m)^{1/n}]$
with $\tau_m$ being the maximal eigenvalue of 
$\mathcal{T}=\prod_{s=1}^n(\partial\mathbf{X}_{s+1}/\partial\mathbf{X}_s)$.
The variation of $\Lambda_m$ as a function of 
$\gamma$ is shown in Fig.~\ref{QCchaosKT}(b4).
It is obvious that $\Lambda_m$ remains zero for $\gamma\lesssim2$, 
due to the regularity of the model.
However, it grows with increasing $\gamma$ as soon as $\gamma>2$,
indicating the onset of chaos in the model. 
Here, it is worth mentioning that even though the transition 
to chaos in the KT model depends on the value of $\beta$ \cite{QianWR2021}, 
the main results of this work are independent of it.
This has been carefully checked in our studies.
We thus fixed $\beta=\pi/3$ in the present work.

By comparing Fig.~\ref{QCchaosKT}(a4) with \ref{QCchaosKT}(b4), 
we note that the upturns in $\widetilde{\la r\ra}$ and $\Lambda_m$ 
are in agreement with each other, suggesting a good quantum-classical correspondence.
This is however expected in the infinite time limit, 
whether it still holds for finite time case remains unknown.
In the rest of this section, we address this question in the KT model
by exploring how the classical chaotic measure, 
defined through the finite time trajectories, 
relates to the degree of quantum chaos.

 \begin{figure}
  \includegraphics[width=\textwidth]{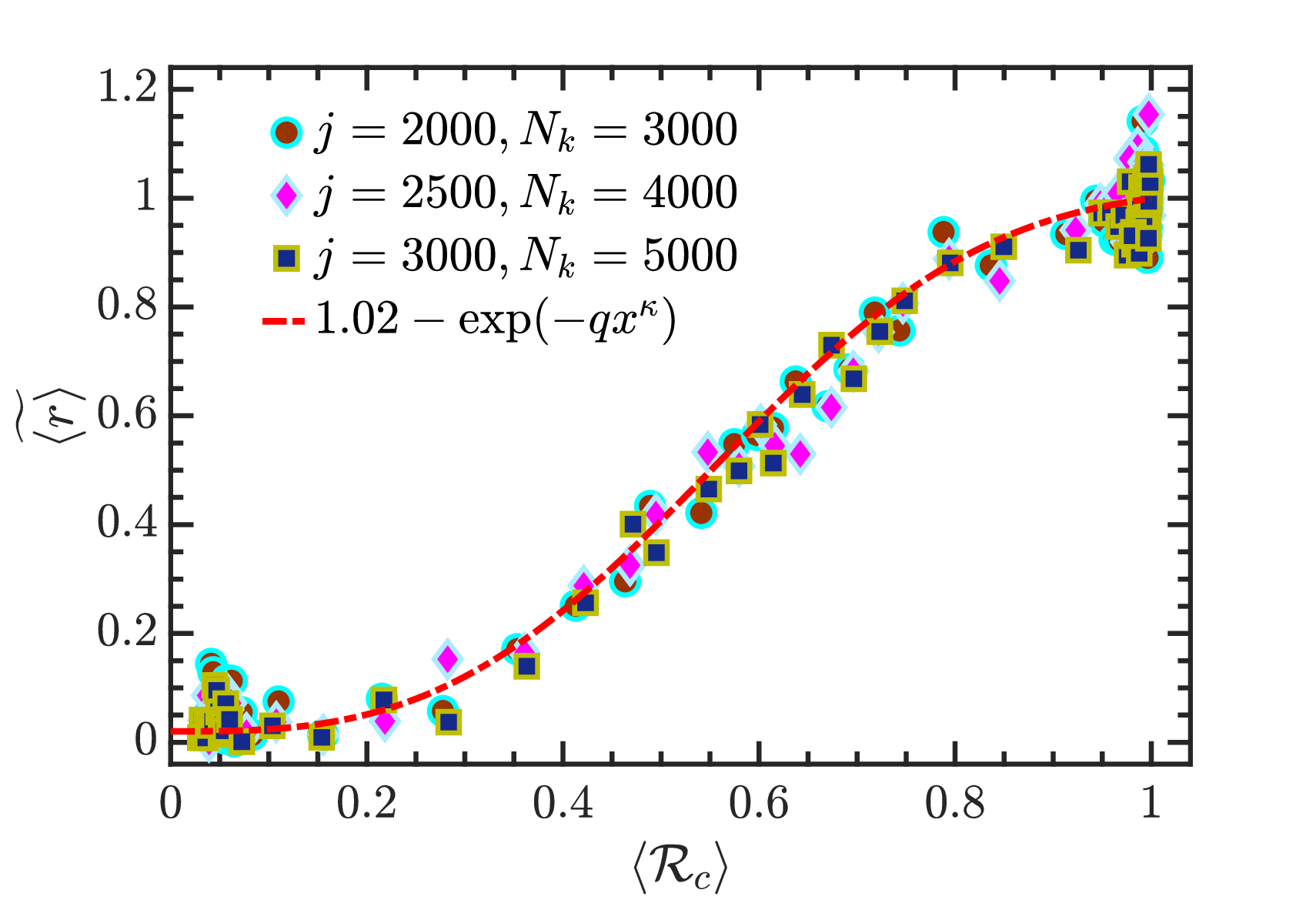}
  \caption{Quantum chaotic measure $\widetilde{\la r\ra}$ in (\ref{Ravr})
  as a function of the finite time classical chaotic measure $\la\mathcal{R}_c\ra$ in (\ref{AvRc})
  for the KT model with several values of $j$ and $N_k$.
  Here, $\la\mathcal{R}_c\ra$ is calculated by averaging over 
  an ensemble with $1600$ initial conditions. 
  The red dot-dashed curve denotes the best fitting of the data and it has the
  form $y=1.02-\exp(-qx^\kappa)$ with $\kappa\approx2.9892$ and $q\approx3.8834$.
  Other parameter: $\beta=\pi/3$.}
  \label{QvsC}
 \end{figure}

\subsection{Dynamical chaos in KT model}
    
The aim of our studying requires us to  
quantify the degree of chaos for a given finite time trajectory.
To this end, we focus on a classical trajectory evolved for $N_k$ kicks.
As a result, a finite time trajectory is described by $N_k$ 
points in the classical phase space.    
The distribution of these $N_k$ points in the phase space is strongly
dependent on whether the classical dynamics is regular or chaotic. 
For regular dynamics, $N_k$ points should exhibit clustering  
and distribute with certain structures in the phase space. 
In contrast, the chaotic dynamics results in $N_k$ points 
being randomly scattered, leading to a structureless phase space.

To quantitatively capture above features exhibited by a finite time classical trajectory, 
we divide the phase space into $M_c$ cells with equal size, 
and consider the number of occupied cells $M_{p}$. 
Obviously, too small and/or too large value of $M_c$ are not allowed, 
as they will wipe out the difference between the regular and chaotic dynamics.  
The suitable value of $M_c$ can be determined via 
the approach proposed by Casati and coworkers \cite{HCasati2024}.
 According to their method, we need to define the Shannon entropy for a certain cell.
 As an arbitrary cell is either occupied or not occupied, 
 the Shannon entropy of a cell can be defined as 
 \be
    S_e(M_c)=-p\ln p-(1-p)\ln(1-p).
 \ee 
 Here, $p=1-(1-1/M_c)^{N_k}$ is the occupation probability of our considered cell
 for $N_k$ points that are randomly distributed over $M_c$ cells.
 Then, by maximizing the Shannon entropy with respect to $M_c$, 
 one can find that the optimal value of $M_c$ is given by
 $M_c=N_k/\ln 2\approx N_k$. 
 In the numerical simulation, one can take $M_c=N_k$ and define the measure of chaoticity   
 of a finite time trajectory as 
 \be \label{KTRc}
    \mathcal{R}_c=\frac{M_r}{N_k},
 \ee
 where $M_r=M_p/(pM_c)N_k$ is the normalized $M_p$.
 One can expect that $\mathcal{R}_c$ will have vanishingly small 
 values for regular trajectories, while it has the order of 
 magnitude $O(1)$ for the chaotic trajectories.    
 
 We find that $\mathcal{R}_c$ in (\ref{KTRc}) indeed enables us to quantify the 
 degree of chaos of the finite time trajectories, as confirmed by its 
 probability distribution for an ensemble of initial conditions, defined by 
 \be
   P_{KT}(\mathcal{R}_c)=\frac{1}{\mathcal{N}_{s}}\sum_{u=1}^{\mathcal{N}_{s}}
         \delta(\mathcal{R}_c-\mathcal{R}_{c,u}),
 \ee 
 where $\mathcal{N}_s$ is the number of initial conditions in the given ensemble.
 The regular behavior for the classical trajectories gives rise to 
 vanishingly small values of $\mathcal{R}_c$, as 
 both the average and fluctuation of $P_{KT}(\mathcal{R}_c)$ are tiny, 
 while the chaotic trajectories would result in a narrow 
 distribution $P_{KT}(\mathcal{R}_c)$ around $\mathcal{R}_c\simeq1$.
    
 In Fig.~\ref{PdfKT}, we show how $P_{KT}(\mathcal{R}_c)$ varies for 
 several values of $\gamma$ and $N_k$.
 On the one hand, as seen in the first column of Fig.~\ref{PdfKT},
 $P_{KT}(\mathcal{R}_c)$ has small average value and width in the regular regime, 
 regardless of the number of kicks.
 Moreover, we see that the average and width of 
 $P_{KT}(\mathcal{R}_c)$ decrease with increasing $N_k$.
 We thus expect that $P_{KT}(\mathcal{R}_c)$ will become a 
 delta distribution located at $\mathcal{R}_c=0$ in the limit $N_k\to\infty$.
 On the other hand, in the chaotic case, the distribution $P_{KT}(\mathcal{R}_c)$
 exhibits a concentration with an average value close to $1$,
 as demonstrated in the last column of Fig.~\ref{PdfKT}.
 Similar to the regular case, the width of $P_{KT}(\mathcal{R}_c)$ for the 
 chaotic trajectories also decreases when we increase $N_k$. 
 In the limit $N_k\to\infty$, one can expect that $P_{KT}(\mathcal{R}_c)$ 
 will take the form $P_{KT}(\mathcal{R}_c)=\delta(\mathcal{R}_c-1)$.
 For the mixed regime, as shown for the case of $\gamma=2.3$ 
 in the second column of Fig.~\ref{PdfKT},
 the distribution $P_{KT}(\mathcal{R}_c)$ is characterized by double peak shape.
 Two peaks in $P_{KT}(\mathcal{R}_c)$ are, respectively, corresponding to 
 the regular and chaotic trajectories. 
 We further note that the smoothness of $P_{KT}(\mathcal{R}_c)$ 
 is enhanced by increasing $N_k$.

 \begin{figure}
  \includegraphics[width=\textwidth]{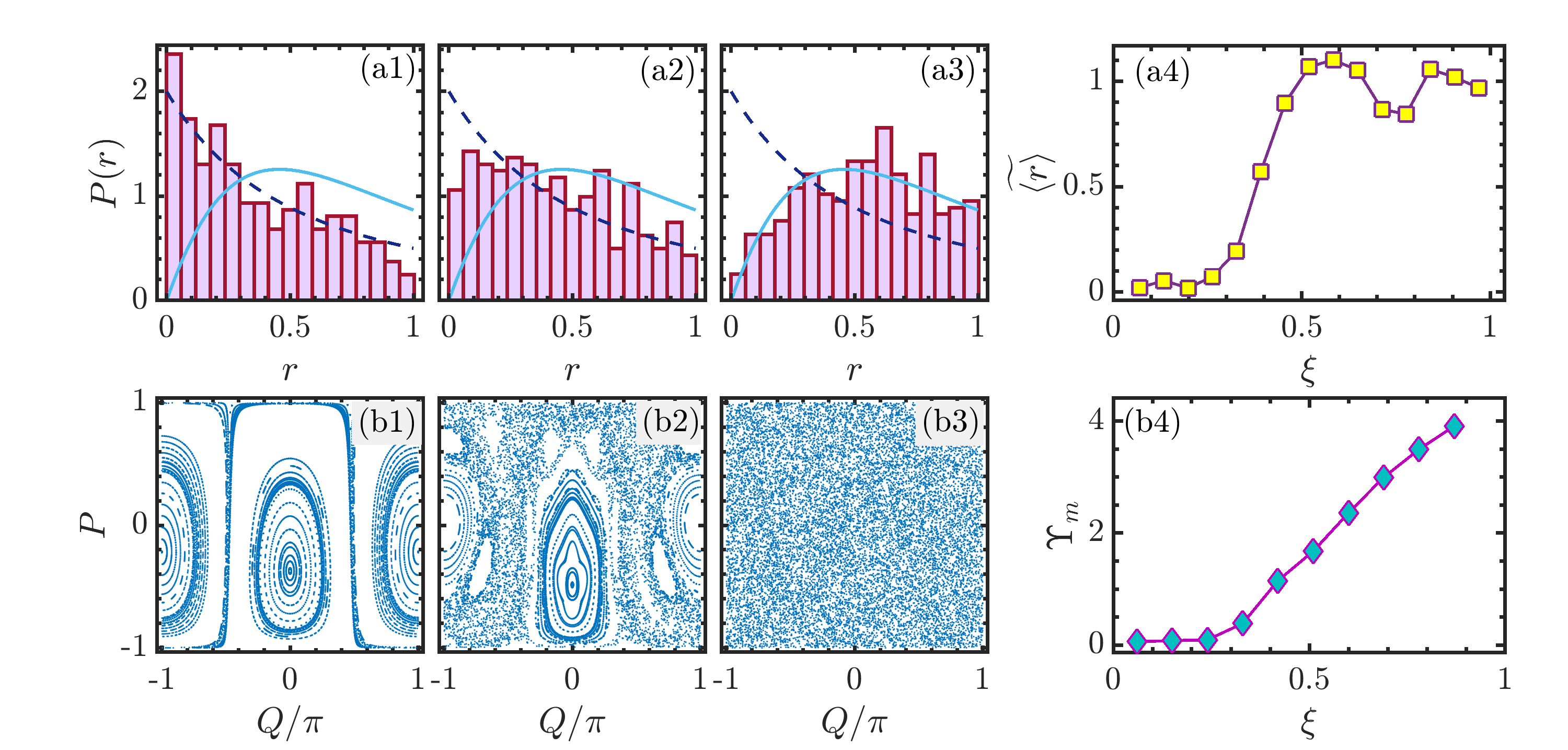}
  \caption{(a1)-(a3): Probability distribution, $P(r)$, of the level spacing ratio (\ref{GapR}) 
  of the Dicke model for $\xi=0.1$ (a1),
  $\xi=0.3$ (a2), and $\xi=1$ (a3) and energy levels $E_k\in[1.05,1.22]$ with $N=80$.
  The dashed and solid curves in each panel denote $P_P(r)$ 
  and $P_{WD}$ in (\ref{PWD}), respectively.
  (a4) Rescaled averaged ratio $\widetilde{\la r\ra}$ in (\ref{Ravr}) of the Dicke model 
  for the energy levels $E_k\in[1.05,1.22]$ as a function of $\xi$ with $N=80$. 
  (b1)-(b3): Classical Poincar\'e sections in $(P,Q)$ plane of the Dicke model
  for $81$ random initial conditions with $\xi=0.1, 0.3$, 
  and $\xi=1$ [from (b1) to (b3)] and a fixed energy $\mathcal{E}=1.2$.
  Each initial condition has been evolved for $t=3000$. 
  (b4) Phase space averaged Lyapunov exponent $\Upsilon_m$ (\ref{AvgLp})
  of the Dicke model as a function of $\xi$.
  Here, $\Upsilon_m$ is numerically obtained 
  by averaging over $5000$ initial conditions, 
  each evolved for $t=1\times10^3$. 
  Other parameters: $\omega=\omega_0=1$.}
  \label{ChaosDicke}
 \end{figure}

 Above features of $P_{KT}(\mathcal{R}_c)$ justify the correctness of
 $\mathcal{R}_c$ for measuring the chaoticity 
 of the finite time classical trajectories. 
 They also imply that the degree of chaos for classical KT model 
 in a finite time can be quantified by the average of $\mathcal{R}_c$, defined by
 \be \label{AvRc}
   \la\mathcal{R}_c\ra=\sum_u \mathcal{R}_{c,u}P_{KT}(\mathcal{R}_c).
 \ee
 To see whether the quantum-classical correspondence still holds in finite time,
 we examine the dependence between $\la\mathcal{R}_c\ra$ 
 and the measure of quantum chaos, given by $\widetilde{\la r\ra}$ in Eq.~(\ref{Ravr}).
 In Fig.~\ref{QvsC}, we show how $\widetilde{\la r\ra}$ correlates with $\la\mathcal{R}_c\ra$
 for different values of $j$ and $N_k$. 
 The collapse of data is clearly visible, indicating that the behavior 
 of $\widetilde{\la r\ra}$ is in good agreement with $\la\mathcal{R}_c\ra$. 
 In particular, a numerical fitting shows that the variation of $\widetilde{\la r\ra}$
 with $\la\mathcal{R}_c\ra$ can be well captured by a function of the form
 \be
  y=1.02-\exp(-qx^\kappa), 
 \ee
 which has been plotted as the red dot-dashed curve in Fig.~(\ref{QvsC}). 
 For the KT model, we have found that $\kappa\simeq2.9892$ and $q\simeq3.8834$ are independent 
 of the values of $j$ and $N_k$. 
 
 These results not only verify the usefulness of $\mathcal{R}_c$ for measuring the chaoticity 
 of the finite time trajectories, but also reveal that the 
 quantum-classical correspondence still holds for the finite time case.
 However, as the KT model is a single particle system, further investigation 
 of quantum many-body systems is required in order to strengthen above statement. 
 We thus proceed to perform an analysis in the celebrated Dicke model.

 \section{Dicke model} \label{third}
  
 The Dicke model consists of an ensemble of $N$ spin-$1/2$ atoms
 interacting with a single bosonic mode \cite{Dicke1954,Garraway2011} and plays a 
 fundamental role in a wide range of fields, including various phase transitions
 \cite{WangYK1973,Vidal2006,Bastidas2012,Bakemeier2012,Castanos2012,Brandes2013,
 Perez2017,Gietka2021,Lewis2021,Corps2021,DasP2022,DasP2023,Lobez2016,Emary2003},
 quantum chaos and thermalization 
 \cite{Emary2003,EClive2023,Altland2012,HouX2004,Magnani2015,Magnani2016,
 Carlos2016,Buijsman2017,Lerma2019,QianWR2020,Villasenor2020,
 Villasenor2023,Kirkova2023,Tiwari2023}, 
 and information scrambling \cite{Alavirad2019,LewisS2019}.
 It has also been applied to understand several critical phenomena
 observed in different experimental platforms \cite{Baumann2010,ZhangZ2018,Kirton2019}. 
 A very recent review on the properties and applications 
 of the Dicke model can be found in Ref.~\cite{Villasenor2024}.
   
 The Hamiltonian of the Dicke model can be written as
 \be \label{DickeH}
   H_D=\omega a^\dag a+\omega_0J_z+\frac{2\xi}{\sqrt{N}}(a+a^\dag)J_x, 
 \ee
 where $\omega$, $\omega_0$, and $\xi$ correspond to the frequency of the bosonic mode, 
 energy splitting of atoms, and the strength of atom-field interaction.
 The bosonic annihilation (creation) operator is denoted by $a$ ($a^\dag$), while
 $J_a (a=x,y,z)$ are the collective spin-$1/2$ operators along $a$ axis 
 and satisfy the $\mathrm{SU}(2)$ commutation relations.
 
 As the Hamiltonian $H_D$ (\ref{DickeH}) commutes with $\mathbf{J}^2=\sum_a J_a^2$, 
 one can divide the Hilbert space into different subspaces 
 distinguished by the eigenvalues of $\mathbf{J}^2$, given by $j(j+1)$.
 We restrict ourselves in the subspace with $j=N/2$ and the dimension of the 
 Hilbert space is $\mathcal{D}=(N+1)(\mathcal{N}_{tr}+1)$ with
 $\mathcal{N}_{tr}$ being the truncation number of the bosonic mode.
 Moreover, the conservation of the parity $\Pi=(-1)^{j+J_z+a^\dag a}$ for $H_D$
 enables us to further separate the Hilbert space into even- and odd-parity subspaces. 
 In this work, we focus on the even-parity subspace, so that 
 the dimension of the Hilbert is given by 
 $\mathcal{D}_e=(N/2+1)(\mathcal{N}_{tr}+1)-\mathcal{N}_{tr}/2$ for even $N$.
 The requirement of the convergence of the numerical 
 results leads us to take $\mathcal{N}_{tr}=380$. 
 We have carefully checked that our results 
 are unchanged for increasing $\mathcal{N}_{tr}$. 
 
 Several interesting features have been found in the Dicke model.
 Among them, the transition to chaos is an intriguing topic 
 and has been triggering many endeavors to explore it, 
 from both static \cite{Emary2003,Magnani2015,Carlos2016,Magnani2016,Magnani2014,Magnani2024} 
 and dynamical aspects \cite{Lerma2019,LewisS2019,Furuya1998,SongL2009,Utso2014,Carlos2019}. 
 Generally speaking, the spectral statistics of $H_D$ turns from the Possion statistics
 to the Wigner-Dyson statistics around $\xi_c\simeq\sqrt{\omega_0\omega}/2$. 
 However, a detailed study reveals that the precise value of $\xi_c$ 
 also strongly depends on the energy \cite{Carlos2016,QianWR2020,QianWR2024}.

 In Figs.~\ref{ChaosDicke}(a1)-\ref{ChaosDicke}(a3), we show how the 
 distribution of level spacing ratio $r_k$ in (\ref{GapR}), 
 $P(r)$, varies with increasing $\xi$
 for the energies $E_k\in[\bar{E}-0.15,\bar{E}+0.02]$ 
 with $\bar{E}=1.2$ and the system size $N=80$. 
 We see that $P(r)$ undergoes an obvious crossover from
 the Possion distribution to the Wigner-Dyson distribution 
 as $\xi$ is increased, suggesting the onset of 
 chaos at larger values of $\xi$.
 Moreover, as we focus on the energy levels with higher energies,
 $P(r)$ displays a visible deviation from $P_P(r)$ 
 at $\xi<\sqrt{\omega_0\omega}/2$, as observed in Fig.~\ref{ChaosDicke}(a2). 
 The transition to chaos with increasing $\xi$ is more clearly revealed by 
 the upturn in the rescaled averaged level spacing ratio $\widetilde{\la r\ra}$ 
 around $\xi\sim0.2$, as seen in Fig.~\ref{ChaosDicke}(a4).

 \begin{figure}
  \includegraphics[width=\textwidth]{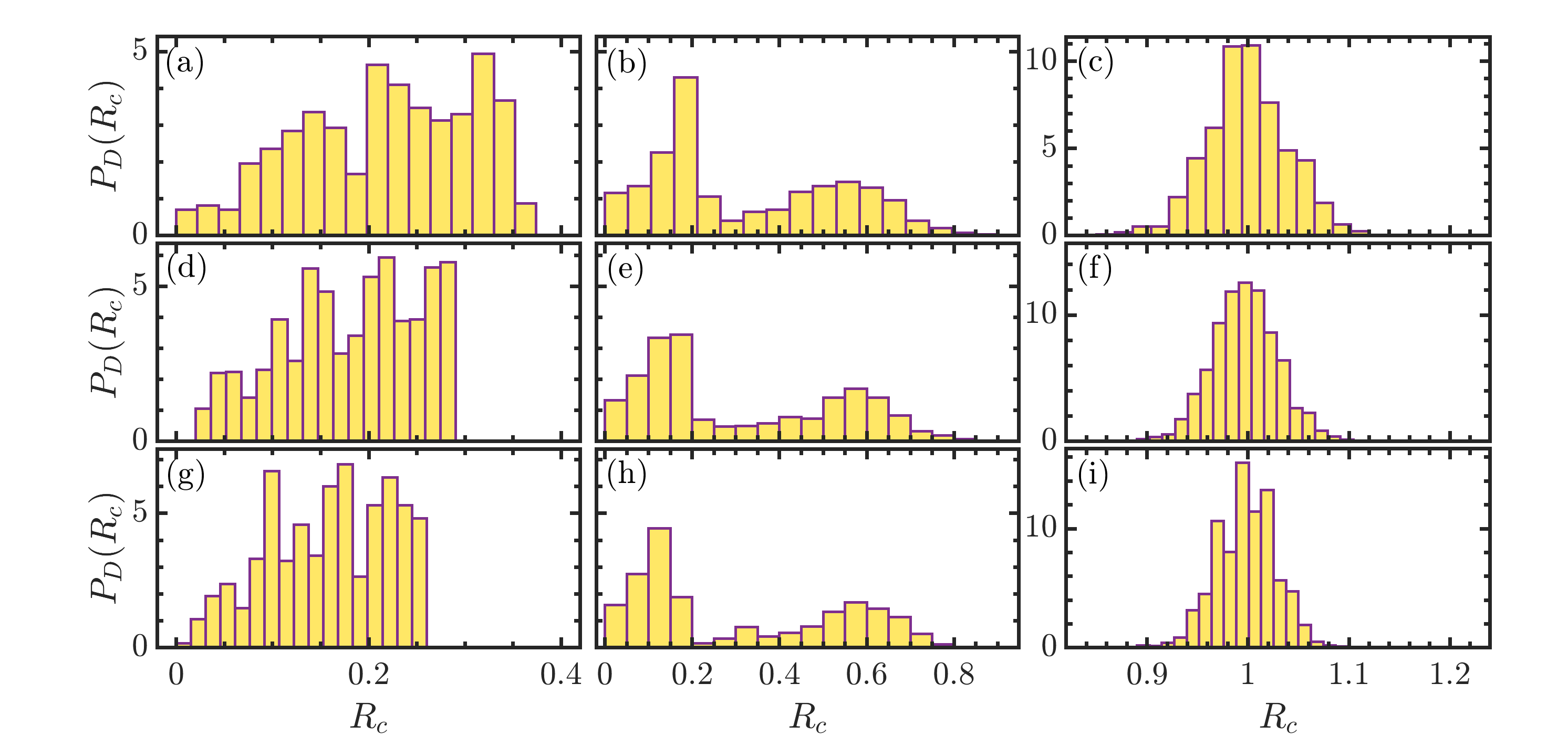}
  \caption{(a)-(c): Probability distribution, $P_{D}(R_c)$, of $R_c$ in Eq.~(\ref{DRc}) 
  for $\xi=0.1$ (a), $\xi=0.3$ (b) and $\xi=1$ (c) with $T_m=1000$.
  (d)-(f): $P_{D}({R}_c)$ for the same values of $\xi$ as in panels (a)-(c) with $T_m=1500$. 
  (g)-(i): $P_{D}({R}_c)$ for the same $\xi$ values as in panels (a)-(c) with $T_m=2000$. 
  Here, $T_m$ denotes the evolution time of our considered trajectories and
  an ensemble of $1600$ initial conditions has been employed to calculate 
  the probability distribution $P_{D}(R_c)$.
  Other parameters: $\omega=\omega_0=1$ and the traversal time $T_r\simeq 6.2$.}
  \label{PdfDicke}
 \end{figure}

 As in the KT model, the presence of chaos in $H_D$ can be understood as
 a quantum analogue of the classical chaos.
 The classical counterpart of the Dicke model is obtained by calculating 
 the expectation value of $H_D$ in the tensor product state 
 $|CS\ra=|\alpha\ra\otimes|z\ra$ \cite{Aguiar1992}.
 Here, $|\alpha\ra$ and $|z\ra$ are the Glauber and 
 Bloch coherent states, defined by \cite{ZhangW1999}
 \be
   |\alpha\ra=e^{-|\alpha|^2}e^{\alpha a^\dag}|0\ra,\quad
   |z\ra=\frac{1}{(1+|z|^2)^j}e^{zJ_+}|j,-j\ra,
 \ee
 where $\alpha,z\in\mathbb{C}$, $|0\ra$ is the vacuum state of the bosonic mode,
 $J_+=J_x+iJ_y$, and $|j,-j\ra$ fulfills $J_z|j,-j\ra=-j|j,-j\ra$. 
 We parametrize $\alpha, z$ as \cite{Magnani2015,Magnani2016}
 \be 
    \alpha=\sqrt{\frac{j}{2}}(q+ip),\quad 
    z=\sqrt{\frac{1+P}{1-P}}e^{iQ},
 \ee
 where $(p,q)\in\mathbb{R}^2$ are the canonical 
 variables of the bosonic sector, while $(P,Q)\in\mathbb{R}^2$
 with $-1\leq P\leq1, -\pi\leq Q\leq\pi$ represent
 the canonical variables of the atomic sector. 
 Then, the classical Hamiltonian of the Dicke model can be written as
 \be \label{CDickeH}
    \mathcal{H}_{D}^c=\frac{\la CS|H_D|CS\ra}{j}
         =\omega_0P+\frac{\omega}{2}(p^2+q^2)+2\xi q\cos Q\sqrt{1-P^2}.
 \ee 
 Consequently, the classical equations of motion (CEM) are given by
 \begin{align} \label{CEqM}
   &\dot{q}=\omega p,\ 
   \dot{p}=-\omega q-2\xi\cos Q\sqrt{1-P^2}, \notag \\
   &\dot{Q}=\omega_0-\frac{2\xi qP\cos Q}{\sqrt{1-P^2}},\ 
   \dot{P}=2\xi q\sin Q\sqrt{1-P^2}.
 \end{align}
with the initial condition $(p_0,q_0,P_0,Q_0)$.  

The chaos presented by $\mathcal{H}_D^c$ in (\ref{CDickeH}) 
with increasing $\xi$ can be uncovered using the Poincar\'e section.
To define the Poincar\'e section at certain energy $\mathcal{E}$, 
we consider a plane with $p=0$.
As the energy of the system is conserved, i.~e. $\mathcal{H}_D^c(p=0,q,P,Q)=\mathcal{E}$,
this plane is actually characterized by the variables $(P,Q)$.
The Poincar\'e section is then defined as the intersection of the 
classical trajectories with this surface.

In Figs.~\ref{ChaosDicke}(b1)-\ref{ChaosDicke}(b3), we plot Poincar\'e 
sections for different values of $\xi$ with energy $\mathcal{E}=1.2$.  
For smaller values of $\xi$, such as the one shown in Fig.~\ref{ChaosDicke}(b1),
the Poincar\'e section is dominated by regular structure,
indicating the regular dynamics in the classical Dicke model. 
This is in good agreement with quantum case in Fig.~\ref{ChaosDicke}(a1).
When the value of $\xi$ increases the chaotic trajectories 
appear, leading to a mixed Poincar\'e section in which the
regular islands are coexisting with a chaotic sea, 
as exemplified in Fig.~\ref{ChaosDicke}(b2) for the case of $\xi=2.3$. 
As $\xi$ is further increased, e.~g. $\xi=1$ case depicted in Fig.~\ref{ChaosDicke}(b3), 
the Poincar\'e section is fully covered by chaotic trajectories, 
although there may still exist several invisible regular islands.

To quantify the chaotic behavior in the Dicke model, we consider the 
phase space averaged Lyapunov exponent, defined by
\be
   \Upsilon_m=\int u_m(P,Q)d\mathcal{A},
\ee
where $d\mathcal{A}=dPdQ$ is the area element of the phase space.
Here, $u_m(P,Q)$ is the largest Lyapunov exponent and can be  
calculated through the numerical approach in Refs.~\cite{Carlos2016,QianWR2024}.   
In Fig.~\ref{ChaosDicke}(b4), we plot $\Upsilon_m$ as 
a function of $\xi$ with $\mathcal{E}=1.2$.
For small values of $\xi$, $\Upsilon_m$ stays vanishingly small
until $\xi\sim0.2$, from which it starts to grow as $\xi$ is increased,
indicating the onset of chaos.
It is worth pointing out that the behavior of $\widetilde{\la r\ra}$ 
is in good agreement with $\Upsilon_m$.

 \begin{figure}
  \includegraphics[width=\textwidth]{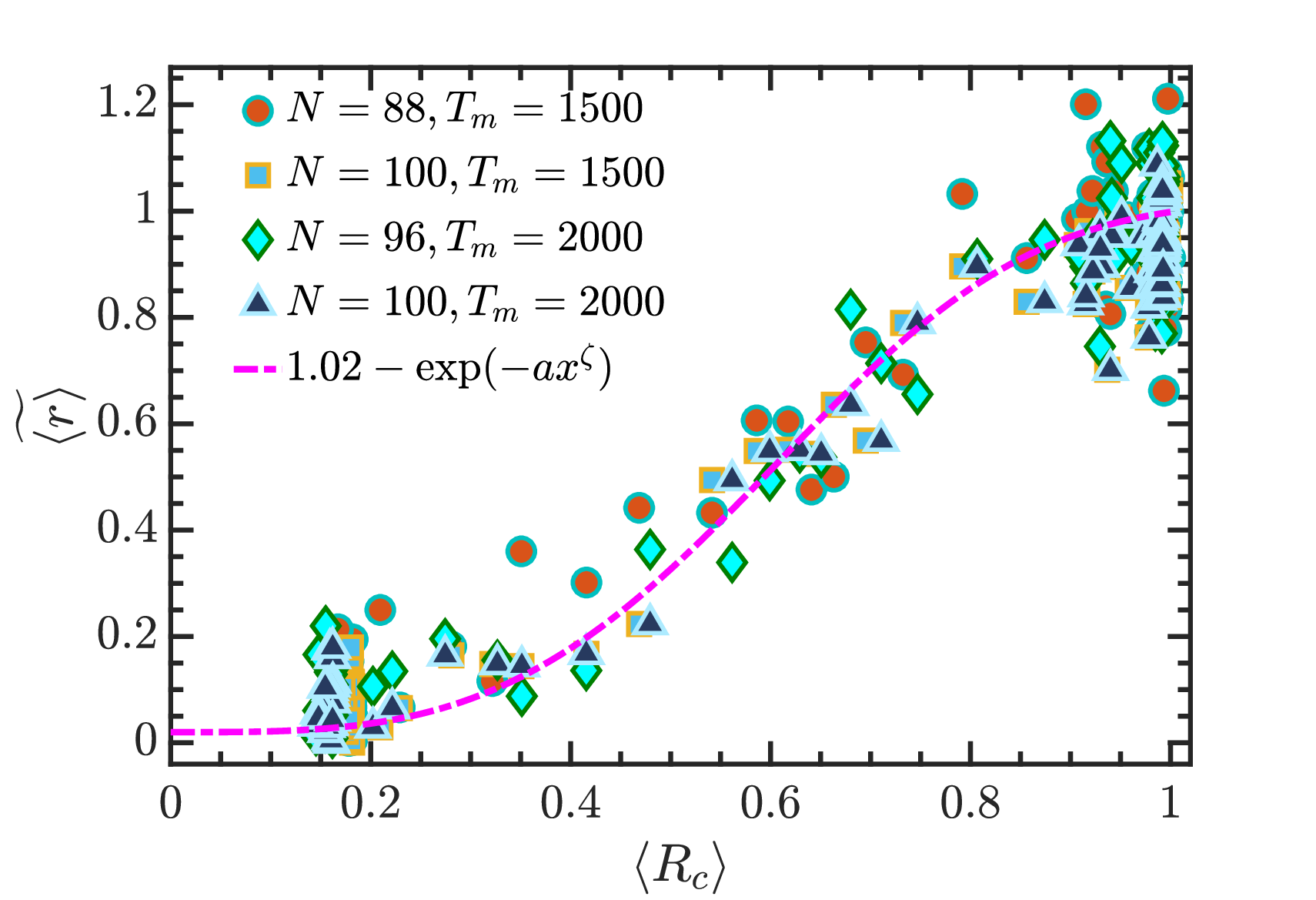}
  \caption{Quantum chaotic measure $\widetilde{\la r\ra}$ in (\ref{Ravr})
  as a function of the finite time classical chaotic measure $R_c$ in (\ref{DAvRc})
  for the Dicke model with several values of the system sizes $N$ and $T_m$.
  Here, $T_m$ is the evolution time of the classical trajectories and
  $R_c$ is calculated by averaging over an ensemble with $1600$ initial conditions. 
  Moreover, $\widetilde{\la r\ra}$ are calculated for the energy levels that satisfy 
  $E\in[\bar{E}-0.15,\bar{E}+0.02]$ with $\bar{E}=1.2$.
  The magenta dot-dashed curve denotes the best fitting of the data and it has the
  form $y=1.02-\exp(-ax^\zeta)$ with $\zeta\approx3.3885$ and $a\approx3.8328$.
  Other parameters: $\omega=\omega_0=1$.}
  \label{QvsCDicke}
 \end{figure}

Different from the KT model, the classical dynamics of the 
Dicke model depends on the energy of the model.
This requires us to consider the energy shell in which the 
underlying classical dynamics should not be changed. 
The energy shell studied in this work is defined by 
$E_k\in[\bar{E}-0.15,\bar{E}+0.02]$ with $\bar{E}=1.2$.
We have checked that there are no substantial variations 
in the corresponding classical dynamics for this energy interval.
Moreover, we would also like to point out that our main results 
still hold for other choice of energy shells, as long as 
the classical dynamics keeps unchanged in the chosen energy shell. 
     
Here, we see again a good correspondence between the classical 
and quantum chaotic features in the Dicke model.
As we have pointed out in the KT model, this correspondence 
is expected in the infinite time limit. 
In order to get more insights into the correspondence principle
in the studies of quantum chaos, it is also necessary to 
explore the connection between the quantum and classical 
chaotic properties in the finite time.

\subsection{Dynamical chaos in Dicke model}

As we have done in the KT model, the chaoticity of the finite time trajectories in the Dicke model 
is also measured through the characters of the points in the Poincar\'e section.  
However, different from the KT model, the classical dynamics of the Dicke model is
governed by the differential equations, rather than the mapping. 
Hence, it is required to determine how many points of a classical 
trajectory evolved up to time $T_m$ are taken in the Poincar\'e section. 
By defining the time between two successive 
crossings of the Poincar\'e section as the traversal time $T_r$, 
the number of points in the Poincar\'e section 
for a trajectory with $T_m$ iterations is $N_m=T_m/T_r$.

Similar to the KT model, the distribution features of these $N_m$ 
points in the Poincar\'e section depend on the chaoticity of the trajectory.
To analyze the chaotic behavior of a finite time trajectory, 
we also discretize the Poincar\'e section into a grid of $M_d$ cells,
and count how many cells are occupied for a given trajectory.
This means that there exists an optimal value of $M_d$.  
Following the same discussion as in the KT model,
it is easy to show that the best choice of $M_d$ 
is given by $M_d=N_m/\ln 2\simeq N_m$.  
We thus define the chaotic measure of a finite time trajectory as
\be \label{DRc}
   R_c=\frac{N_R}{N_m},
\ee
where $N_R=M_R/(pM_d)N_m$ with $M_R$ being the 
number of occupied cells and $p=1-(1-1/M_d)^{N_m}$.

Let us first focus on the distribution of $R_c$ for an 
ensemble of initial conditions, defined by 
\be
    P_D(R_c)=\frac{1}{\mathcal{M}_s}\sum_{v=1}^{\mathcal{M}_s}\delta(R_c-R_{c,v}),
\ee
where $\mathcal{M}_s$ denotes the number of initial conditions in an ensemble.
The variation of $P_D(R_c)$ for an ensemble of $1600$ initial conditions with 
different values of $\xi$ and $T_m$ has been illustrated in Fig.~\ref{PdfDicke}. 
An overall similarities between Fig.~\ref{PdfKT} and Fig.~\ref{PdfDicke} can be clearly identified.
Specifically, since the points in the Poincar\'e section are clustered for the trajectories 
with regular dynamics, the distribution $P_D(R_c)$ is supported over the smaller values of $R_c$ 
and its width decreases with increasing $T_m$, as seen in the first column of Fig.~\ref{PdfDicke}. 
This behavior of $P_D(R_c)$ leads us to expect that it would turn to 
$P_D(R_c)=\delta(R_c)$ as $T_m$ goes to infinity, 
and the same trend is also presented in the KT model.
On the contrary, as the trajectories with chaotic dynamics result in randomly scattered points 
in Poincar\'e section, $P_D(R_c)$ is well described by a narrow 
distribution around $R_c\simeq1$, as shown in the last column of Fig.~\ref{PdfDicke}. 
We further note that the distribution of $R_c$ for the chaotic trajectories 
becomes more and more narrow as $T_m$ is increased, suggesting
$P_D(R_c)$ approaches a delta distribution centered 
at $R_c=1$ in the $T_m\to\infty$ limit, as exhibited by the KT model.         
The coexistence of the regular and chaotic trajectories in the mixed regime
gives rise to the double peak shape of $P_D(R_c)$, as demonstrated 
in the middle column of Fig.~\ref{PdfDicke}.
As observed in the KT model, the smoothness of $P_D(R_c)$ 
can be improved by enhancing $T_m$.

After verifying the ability of $R_c$ in (\ref{DRc}) to capture the 
chaotic degree of the finite time trajectories, we continue to  
analyze the validity of the quantum-classical correspondence in
the Dicke model for finite time trajectories.
According to the results in the KT model, the 
chaoticity of the classical Dicke model in finite time 
can be measured by averaging $R_c$ over an ensemble,
\be \label{DAvRc}
   \la R_c\ra=\sum_{v=1}^{\mathcal{M}_s} R_{c,v}P_D(R_c),
\ee
while the degree of quantum chaos is quantified through 
$\widetilde{\la r\ra}$ [cf.~Eq.~(\ref{Ravr})] in a given energy shell. 
The correspondence between the quantum 
and classical chaoticity measures indicates that
the variation of $\widetilde{\la r\ra}$ should 
depend strongly on $\la R_c\ra$.

The variation of $\widetilde{\la r\ra}$ as a function of $\la R_c\ra$
for several system sizes $N$ and evolution times $T_m$ in the Dicke model
is shown in Fig.~\ref{QvsCDicke}.  
The resemblance between Figs.~\ref{QvsC} 
and \ref{QvsCDicke} is clearly visible.
However, as the system size and evolution time that can be approached 
in our numerical simulation of the Dicke model 
are much smaller than in the KT model, the results for the Dicke model
have larger fluctuations, as compared to Fig.~\ref{QvsC}.
Nevertheless, one can see that the fluctuations can be 
suppressed by increasing $N$ and/or $T_m$. 
In particular, as marked by the magenta dot-dashed curve in Fig.~\ref{QvsCDicke},
the dependence of $\widetilde{\la r\ra}$ on $\la R_c\ra$ 
is captured by the same function as in the KT model,
i.~e. $y=1.02-\exp(-ax^\zeta)$.   
For the Dicke model, we obtain $\zeta\simeq3.3885, a=3.8328$, 
which are remarkably close to the corresponding values in the KT model. 
Almost the same relationship between the quantum and classical chaotic measures
has been observed in two totally different quantum systems. 
This fact not only confirms that the correspondence principle also holds 
for the finite time trajectories, but also leads us to expect that
our main findings may still be valid for other quantum 
many-body systems with well-defined classical limit.

\section{Conclusions} \label{fourth}

In this work, we have studied how to measure the degree of chaos
for finite time trajectories and examined the quantum-classical 
correspondence in two different quantum systems, 
namely kicked top and Dicke models.  
By utilizing the distinct features exhibited by the 
regular and chaotic trajectories in the Poincar\'e section, 
we have shown how to quantify the degree of chaos for a finite time classical trajectory. 
The correctness of our defined finite time chaotic measure has been
verified through a detailed analysis of its probability distribution 
in our considered models. 
We have demonstrated that the transition to chaos leaves a strong imprint 
on the statistical properties of the finite time chaotic measure.  
Identifying the average of level spacing ratio as the measure of quantum chaos, 
we have explored the correspondence between the quantum 
and finite time classical chaoticities in both kicked top and Dicke models. 
Notably, the dependence of quantum chaotic measure on its classical counterpart
can be well described by a simple function, which is independent of the model.

Our findings shed more light on the meaning of quantum-classical correspondence and 
provide a quantitative way to understand how the chaos develops with increasing time.
The analytical expression of the relation between quantum 
and classical chaotic measures is independent of specific model. 
This indicates a general standard for ascertaining the quantum-classical correspondence. 
Moreover, the results of this work not only extend the conclusions 
of previous work \cite{HCasati2024}, but also enhance 
our comprehending of quantum statistical physics. 

A natural extension of the present work is to systematically investigate 
whether the quantum and finite time classical chaotic measures have
the same relationship as in this work also in other quantum systems 
with well-defined classical counterpart. 
We expect that the results of these explorations will align with our conclusions.  
Another interesting question should be the analytical 
explanation of why the quantum chaotic measure exhibits
the same dependence on the classical measure for different models. 
It would also be intriguing to examine how the finite time chaotic measure 
links to other quantum chaotic indicators and particularly the dynamical ones,
such as spectral form factor \cite{Kos2018} 
and Krylov complexity \cite{Parker2019,Balasub2022}.

\acknowledgments  

This work was supported by the Slovenian Research 
and Innovation Agency (ARIS) under the 
Grants Nos.~J1-4387 and P1-0306.

\bibliographystyle{apsrev4-1}


%

\end{document}